\documentclass{PoS}

\usepackage{graphics,wrapfig,times}
\usepackage{epsf}
\usepackage{amsmath,amssymb,latexsym,float,algorithm,algorithmic}

\newcommand{\R}{\text{\fontshape{n}\selectfont I\kern-.42exR}}
\newcommand{\1}{\text{\fontshape{n}\selectfont 1\kern-.56exl}}

\PoS{PoS(LAT2005)099}

\title{Reducing the beta-shift in domain wall fermion simulations}
\ShortTitle{Reducing the beta-shift in domain wall fermion simulations}
                                                                        
\author{\speaker{Alban Allko\c{c}i}
\thanks{The authors participation in the conference and part of this work are funded from the NATO grant EAP.RIG.981410.}\\
        Computer Science Section\\
        Polytechnic University of Tirana\\
        Mother Theresa Square\\
        Tirana - Albania\\
        E-mail: \email{alban@fie.upt.al}\\
}

\author{Artan Bori\c{c}i\\
        Physics Department\\
        University of Tirana\\
        Blvd. King Zog I\\
        Tirana - Albania\\
        E-mail: \email{borici@fshn.edu.al}\\
}
                                                                        
\abstract
{The beta-shift induced from dynamical domain wall quarks leads to
increased roughness of the gauge field, thus reversing the effect
of smoothing from the gauge action improvement. By exploiting the
relation of overlap and domain wall fermions in greater detail,we
propose an algorithm which reduces the beta-shift to the level of
dynamical overlap fermions.}
                                                                        
\FullConference{XXIIIrd International Symposium on Lattice Field Theory\\
                 25-30 July 2005\\
                 Trinity College, Dublin, Ireland}

\begin{document}

\section{Introduction}

Lattice QCD with chiral fermions, although computationally expensive,
is the best formulation of QCD on the lattice.
There are two chiral formulations:
a) Domain wall fermions \cite{Ka92,FuSha95} and
b) overlap fermions \cite{NaNe93}, which are closely related
\cite{Borici_00,Borici_qcdna3_intro}.

In this work we focus on the use of domain wall type
fermions for lattice
QCD simulations. Recent dynamical simulations with such fermions
revealed larger chiral symmetry breaking than expected
\cite{Aoki_et_al04}. It was shown that the effect of
gauge action improvement is effectively canceled by the
dynamical domain wall fermion. Since there is no such problem
with dynamical overlap fermions one wonders if there is something
wrong with domain wall fermion formulation.

But as mentioned above these formulations are closely related and
therefore if anything is to be blamed it is the implementation
of domain wall fermions. It is the purpose of this work to show
that the state-of-the-art of the dynamical domain wall implementation
is not well suited for the state-of-the-art simulation algorithms.
 
\section{Notations}

In this work we will use the truncated overlap fermions
\cite{Borici_TOV} which have
better chiral properties than the standard domain wall fermions
\cite{Brower_et_al_04}.
The corresponding 5-dimensional operator
is given by the $N_5\times N_5$ blocked matrix:
\begin{equation*}
{\mathcal M} =
\begin{pmatrix}
~D_W-\1          & ~~~~~(D_W+\1)P_+ &                & -m(D_W+\1)P_- \\
~~~~~~(D_W+\1)P_-     & D_W-\1      & \ddots         &
  \\
                & \ddots      & \ddots         & ~~~~~~(D_W+\1)P_+
  \\
-m(D_W+\1)P_+ &             & (D_W+\1)P_-    & ~D_W-\1
  \\
\end{pmatrix}
\end{equation*}
where the blocks are matrices defined on the 4-dimensional lattices
using the negative mass Wilson-Dirac operator $D_W$,
$N_5$ is the number of time slices along the fifth Euclidean
dimension and
$P_{\pm}$ are chiral projection operators. Note that
standard domain wall fermions use off-diagonal blocks which omit $D_W$.

Let ${\mathcal M}_1$ be the same matrix as above but with the bare
quark mass $m = 1$. Then it can be shown that
\cite{Borici_00,Borici_qcdna3_intro}:
\begin{equation}\label{det_identity}
\det {\mathcal M}_1^{-1} {\mathcal M} = \det D^{(N_5)}
\end{equation}
where
\begin{equation}
\label{TOV}
D^{(N_5)} = \frac{1 + m}{2} \1 + \frac{1 - m}{2} \gamma_5
        \frac{\1 - T^{N_5}}{\1 + T^{N_5}}
\end{equation}
with $T$ the transfer matrix along the fifth dimension given by:
\begin{equation*}
T = \frac{\1 + H_W}{\1 - H_W}, ~~~~~H_W = \gamma_5 D_W
\end{equation*}
Note also that in the large $N_5$ limit $D^{(N_5)}$
approaches the Neuberger overlap operator \cite{Ne98}:
\begin{equation*}
D = \frac{1 + m}{2} \1 + \frac{1 - m}{2} \gamma_5 \text{sign}(H_W)
\end{equation*}

\section{The problem}

The standard way one deals with the fermion determinant is expressing it
as a Gaussian integral over pseudofermion fields:
\begin{equation*}
|\det D^{(N_5)}|^2 = |\det {\mathcal M}_1^{-1} {\mathcal M}|^2 
= \int ~[d\Phi^*d\Phi] ~e^{~-||{\mathcal M}^{-1} {\mathcal M}_1\Phi||^2}
\end{equation*}
The resulting effective fermion action is given by:
\begin{equation*}
S_{\text{PF}} = ||{\mathcal M}^{-1} {\mathcal M}_1\Phi||^2
\end{equation*}

From equation \ref{det_identity}
we expect the action to have one contribution from
the $D^{(N_5)}$ operator and some extra terms:
\begin{equation*}
S_{\text{PF}} = ||[D^{(N_5)}]^{-1} \chi_1||^2
+ \sum_{i = 2}^{N_5} ||C_i \chi_i||^2
\end{equation*}
where $\chi_i$ are pseudofermion fields and $C_i$ are
four dimensional matrices which in general may depend on gauge
fields.
This form of the action will be explicitly calculated below in this
paper.

We note the extra terms in the action and we ask whether
they would contribute in the generation of gauge fields.
As long as the simulation
algorithm averages over a large ensemble of pseudofermion
fields $\chi_i$ these terms will cancel to give the correct
determinant. However, fermion algorithms typically construct
molecular dynamics trajectories which keep the pseudofermion
field fixed.
This may cause the exploration of gauge field configurations
which ``feel'' the extra dimension through the extra terms.

These extra pseudofermion terms may be regarded as ''artefacts''
of the algorithm which contribute via the renormalised coupling:
\begin{equation*}
\beta \rightarrow c_1 \beta + \Delta \beta
\end{equation*}
As observed by \cite{Aoki_et_al04} the renormalisation
is such that it drives the gauge field toward the Aoki parity 
broken phase which in turn causes the breaking of
the chiral symmetry.

We believe that this effect should go away if the extra terms
are not present in the action. A direct evidence for this is
not provided here, but simulations with overlap fermions clearly
show that such renormalisation effects are absent.

\section{The solution}

First, let us calculate $S_{\text{PF}}$.
Using algebraic manipulations
as in \cite{Borici_qcdna3_intro} 
and the results of the Appendix we get:
\begin{equation*}
S_{\text{PF}} = ||{\mathcal M}^{-1} {\mathcal M}_1\Phi||^2
              = ||{\mathcal T}^{-1} {\mathcal T}_1 P^{T} \Phi||^2
\end{equation*}
where $\mathcal T$ is given by:
\begin{equation*}
\mathcal{T} =
\begin{pmatrix} P_+ - mP_-     & -T &        &          \\
                               & \1 & \ddots &          \\
                               &    & \ddots & -T       \\
                -T(P_- - mP_+) &    &        & \1       \\
\end{pmatrix}
\end{equation*}
and $P$ is the permutation matrix:
\begin{equation*}
\begin{pmatrix} P_+ & P_- &        &     \\
                    & P_+ & \ddots &     \\
                    &     & \ddots & P_- \\
                P_- &     &        & P_+ \\
\end{pmatrix}
\end{equation*}
It is straightforward to show that (see Appendix):
\begin{equation}
\label{result}
{\mathcal T}^{-1} {\mathcal T}_1 =
\begin{pmatrix}
 [D^{(N_5)}]^{-1} & ~~~~~~   & ~~~~~~       & ~~~~~~   \\
 B^{(2)}          & \1 & ~~~~~~       & ~~~~~~   \\
     \vdots       & ~~~~~~   & ~~~~~~       & ~~~~~~   \\
                  & ~~~~~~   & ~~~~~~\ddots & ~~~~~~   \\
 B^{(N_5)}        & ~~~~~~   & ~~~~~~       & \1 \\
\end{pmatrix}
\end{equation}
where
$B^{(i)} = T^{N_5-i+1}\{(P_- - mP_+)[D^{(N_5)}]^{-1} + \gamma_5 \1\},
i = 2, \ldots, N_5$.
This way we get:
\begin{equation*}
S_{\text{PF}} = ||[D^{(N_5)}]^{-1}{\hat \Phi_1}||^2
+ \sum_{i = 2}^{N_5} ||B^{(i)} {\hat \Phi_1} + {\hat \Phi_i}||^2
\end{equation*}
where ${\hat \Phi_1} = P_+ \Phi_1 + P_- \Phi_{N_5}$
and ${\hat \Phi_i} = P_+ \Phi_i + P_- \Phi_{i-1}, ~i=2,\ldots,N_5$.
It is easy to see that the second term of the right hand side
constitutes the bulk of the coupling constant renormalisation.

This result hints to the following solution of the problem:
{\it express the fermion determinant in such a way that there
are no additional terms in the effective fermion action}.
An obvious solution is to define:
\begin{equation*}
S_{\text{PF}} = ||[D^{(N_5)}]^{-1} \phi||^2
\end{equation*}
where $\phi$ is now the usual pseudofermion field defined on the
four dimensional lattice.
The difficulty with this form is that the differentiation
of (\ref{TOV}) w.r.t. gauge field may yield numerically
unstable expressions and ill-conditioned matrices.

A stable implementation may be defined as in the following:
let $\varepsilon_1$ be defined as the blocked unit direction
along the fifth dimension, i.e.,
\begin{equation*}
\varepsilon_1 = (\1, 0, \ldots, 0)^T
\end{equation*}
Using (\ref{result}) we can write for the fermion determinant:
\begin{equation*}
\label{solution}
|\det D^{(N_5)}|^2
 = |\det \varepsilon_1^T {\mathcal T}_1^{-1}
                         {\mathcal T} \varepsilon_1|^2 
 = |\det \varepsilon_1^T P^T {\mathcal M}_1^{-1}
                             {\mathcal M} P \varepsilon_1|^2 
\end{equation*}
This expression can be used as a starting point
to formulate a simulation
algorithm in terms of numerically stable derivatives of
$\mathcal M$-matrices. Hence, a variation of $D^{(N_5)}$
can be computed using the variations of $\mathcal M$ and
${\mathcal M}_1$.

Note that the equation \ref{result} can be used to compute the inverse
of $D^{(N_5)}$. In this case the linear system to be solved is:
\begin{equation}\label{4d_system}
D^{(N_5)} x = b
\end{equation}
Let $y=(x,y_2\ldots,y_{N_5})$
and $z = (b,0\ldots,0)$ be 5-dimensional vectors.
Then, from eq. \ref{result} (see Appendix) we can write:
\begin{equation*}
{\mathcal T} y = {\mathcal T}_1 z
\end{equation*}
Using eq. \ref{mathcal_T} we get:
\begin{equation*}
{\mathcal M} P y = {\mathcal M}_1 P z
\end{equation*}
In this way, one can use the usual solver for the 5-dimensional
matrix and get the solution $x$ of the 4-dimensional system
\ref{4d_system}
form the first block-component of $y$.

\section{Conclusion}

In this paper we have given an explicit calculation of the effective
pseudofermion action that is used in dynamical
domain wall simulations. We note that the extra pseudofermion terms
may cause the current simulation algorithms to produce
gauge fields in the region of the Aoki phase.

To cure this phenomenon we have proposed a simple solution which
can be
easily implemented using the algebraic relations between overlap and
domain wall fermions.

\pagebreak

\section{Appendix: explicit calculation of $S_{\text{PF}}$}

Multiplying $\mathcal M$ from the right by $P$ we obtain:
\begin{equation*}
\gamma_5
\begin{pmatrix}
(H_W - \1)(P_+ - mP_-) & H_W + \1 &        &          \\
                       & H_W - \1 & \ddots &          \\
                       &          & \ddots & H_W + \1 \\
(H_W + \1)(P_- - mP_+) &          &        & H_W - \1 \\
\end{pmatrix}
\end{equation*}
Further, multiplying this result from the left by the inverse
of the diagonal matrix:
\begin{equation*}
H_5 = \gamma_5 ~\text{diag}(H_W - \1, \ldots, H_W - \1)
\end{equation*}
we get:
\begin{equation}\label{mathcal_T}
\mathcal{T} = H_5^{-1} \mathcal{M} P
\end{equation}

Let $X_{ij}, i,j=1, \ldots, N_5$ be the four dimensional blocks of
the inverse of $\mathcal T$.
Then a straightforward calculation gives:
\begin{equation*}
X_{1j} = \frac{\1}{(P_+ - mP_-) - T^{N_5} (P_- - mP_+)} T^{j-1}
     = \frac{\1}{D^{(N_5)}} \gamma_5 \frac{\1}{\1 + T^{N_5}} T^{j-1}
\end{equation*}
for $j = 1, \ldots, N_5$ and
\begin{equation*}
X_{i1} = T^{N_5-i+1} (P_- - m P_+) X_{11}
\end{equation*}
for $i = 2, \ldots, N_5$ and
\begin{equation*}
X_{ij} = T^{N_5-i+1} (P_- - m P_+) X_{1j}
       + \sum_{k = 0}^{N_5-i} \delta_{i+k,j} T^k
\end{equation*}
for $i = 2, \ldots, N_5, ~j = 2, \ldots, N_5$.

\end{document}